\documentclass[twocolumn,amsmath,amssymb,superscriptaddress]{revtex4-1}
\usepackage{upgreek}
\usepackage{graphicx}

\begin{document}

\preprint{QED/123}

\title{Nucleated dewetting in supported ultra-thin liquid films with hydrodynamic slip}

\author{Matthias Lessel}
\thanks{Contributed equally}
\affiliation{Department of Experimental Physics, Saarland
 University, D-66041 Saarbr\"ucken, Germany}
\author{Joshua D. McGraw}
\thanks{Contributed equally}
\affiliation{Department of Experimental Physics, Saarland University, D-66041 Saarbr\"ucken, Germany}
\affiliation{D\'epartement de Physique, Ecole Normale Sup\'erieure/PSL Research University, CNRS, 24 rue Lhomond, 75005 Paris, France}
\author{Oliver B\"aumchen}
\affiliation{Max Planck Institute for Dynamics \& Self-Organization (MPI-DS), D-37077 G\"ottingen, Germany}
\author{Karin Jacobs}
\thanks{Corresponding author: k.jacobs@physik.uni-saarland.de}
\affiliation{Department of Experimental Physics, Saarland
 University, D-66041 Saarbr\"ucken, Germany}

\date{\today}

\begin{abstract}
This study reveals the influence of the surface energy and solid/liquid boundary condition on the breakup mechanism of dewetting ultra-thin polymer films. Using silane self-assembled monolayers, SiO$_2$ substrates are rendered hydrophobic and provide a strong slip rather than a no-slip solid/liquid boundary condition. On undergoing these changes, the thin-film breakup morphology changes dramatically -- from a spinodal mechanism to a breakup which is governed by nucleation and growth. The experiments reveal a dependence of the hole density on film thickness and temperature.  The combination of lowered surface energy and hydrodynamic slip brings the studied system closer to the conditions encountered in bursting unsupported films. As for unsupported polymer films, a critical nucleus size is inferred from a free energy model. This critical nucleus size is supported by the film breakup observed in the experiments using high speed \emph{in situ} atomic force microscopy.
\end{abstract}

\maketitle

From the simplest solid, liquid and vapor phases, to exotic condensed matter systems, equilibrium phases and phase transitions, along with their dynamics, make up an essential part of the fundamental description of many systems~\cite{Baus08TXT,chaikin95TXT}. This statement is also valid for nanometric liquid films on solid substrates. Thin films can be described through an effective interface potential~\cite{Israelachvili1972, Seemann2001, Seemann2005, Thiele2001,Thiele2003, blossey2012, dietrich13TXT}, $\phi(h)$, in which $\phi$ is the energy per unit area, and $h(x,y)$ is the local film height. The free energy is then a functional~\cite{Thiele2001,Thiele2003, blossey2012, dietrich13TXT}, $F[\phi(h), h]$, of the film height distribution. Depending on the effective interface potential, a flat film with $h = h_0$ everywhere, may be in stable equilibrium or susceptible to breakup~\cite{Oron1997,Reiter1992,Jacobs1998,degennes03TXT}. When, as considered in this work, it is energetically preferred for the liquid to phase separate into regions of differing film thickness, the morphology of the phase separated regions gives crucial information on the breakup mechanism. 

Two dewetting mechanisms can be observed in experiments on SiO$_2$ substrates. One is nucleated dewetting~\cite{Reiter1992,Jacobs1998}, which can be subdivided into heterogeneous and homogeneous (or thermal) nucleation. For heterogeneous nucleation, an external stimulus like a local topographical defect or chemical heterogeneity leads to the breakup; thermal nucleation in supported polymer films has only rarely been observed~\cite{Seemann2001,Seemann2005}. The second mechanism, which occurs spontaneously, is called spinodal dewetting~\cite{Vrij1966,Seemann2001,Becker2003,Tsui2006}. In a spinodally dewetting film, capillary waves at the liquid/air interface are amplified by van der Waals (vdW)  interactions between the air/liquid interface and the liquid/substrate interface. When the amplitude of the fastest growing capillary wave reaches the film thickness, holes are created in the film. The dominant rupture mechanism depends on various experimental parameters~\cite{Seemann2005}; namely, the thickness of the liquid film and the substrate properties, including the surface energy and vdW interactions between film/air interface and the full depth profile of the substrate~\cite{Seemann2001, Seemann2005, isf11TXT}. A polystyrene (PS) film on a pure Si substrate is stable, for example. If a SiO$_2$ layer is added, the film becomes unstable to nucleated dewetting or spinodal rupture~\cite{Seemann2001, Seemann2005}. 

Removing the substrate entirely, thereby creating an unsupported film, 
another system that is unstable to rupture can be studied~\cite{debregas95PRL, dalnoki99PRE, roth06JPSB, Croll2010, Rathfon2011}. As with certain supported films, the interfaces on either side of an unsupported film are attractive~\cite{isf11TXT}. These unsupported films are thus in principle susceptible to spontaneous rupture, but are observed to rupture via nucleation only. The nucleated breakup of freestanding films can occur through a defect as was studied by Debregeas and coworkers~\cite{debregas95PRL} and others~\cite{dalnoki99PRE, roth06JPSB}, or may occur via a random nucleation event~\cite{Croll2010,Rathfon2011}. In the random nucleation case, fluctuations of the liquid film lead to the formation of holes. In contrast to spinodal dewetting, vdW interactions are not the dominant driving force for this process. The liquid film can instead be treated as an energy barrier: bending the surface introduces excess area and results in an energetic cost that may be overcome by random thermal fluctuations~\cite{Croll2010}. 

Here, we make a connection between the breakup of unsupported and supported thin films by taking advantage of hydrodynamic slip. Slip refers to the relative motion between molecules in the fluid and the substrate at the solid/liquid interface, and is characterized by a slip length, $b$. In Navier's linear slip model, $b$ is defined through $u|_{z=0} = b\partial_zu|_{z=0}$ where $u$ is the horizontal component of the fluid velocity and $z$ is the vertical coordinate. SiO$_2$ substrates support no slip in dewetting experiments~\cite{Seemann2001,Seemann2005}. Meanwhile, an unsupported film provides a diverging slip length since the air/liquid interface can support no shear stress; a plug flow is observed~\cite{debregas95PRL}. Solid substrates have been shown to support interfacial slip of polymeric liquids if decorated with hydrophobic surface coatings such as self-assembled monolayers (SAMs)~\cite{Fetzer2005,Fetzer2007,gutfreund13PRE} or amorphous fluoropolymer coatings~\cite{Baeumchen2009a}. Slip lengths depend on the molecular weight of the polymer and on the specific coating applied~\cite{Baeumchen2009a}. Thus, using the right combination of polymer and coating, slip lengths in the range of microns for polymer films with thicknesses  $h_0 < 10$\,nm can be achieved. Since here $b/h_0 \gg 1$, we reach a situation that can be assumed as nearly plug flow, thus bridging the gap between unsupported and supported films. 
\begin{figure}[t!]
\centering
\begin{minipage}{\columnwidth}%
	 \includegraphics[width=\columnwidth]{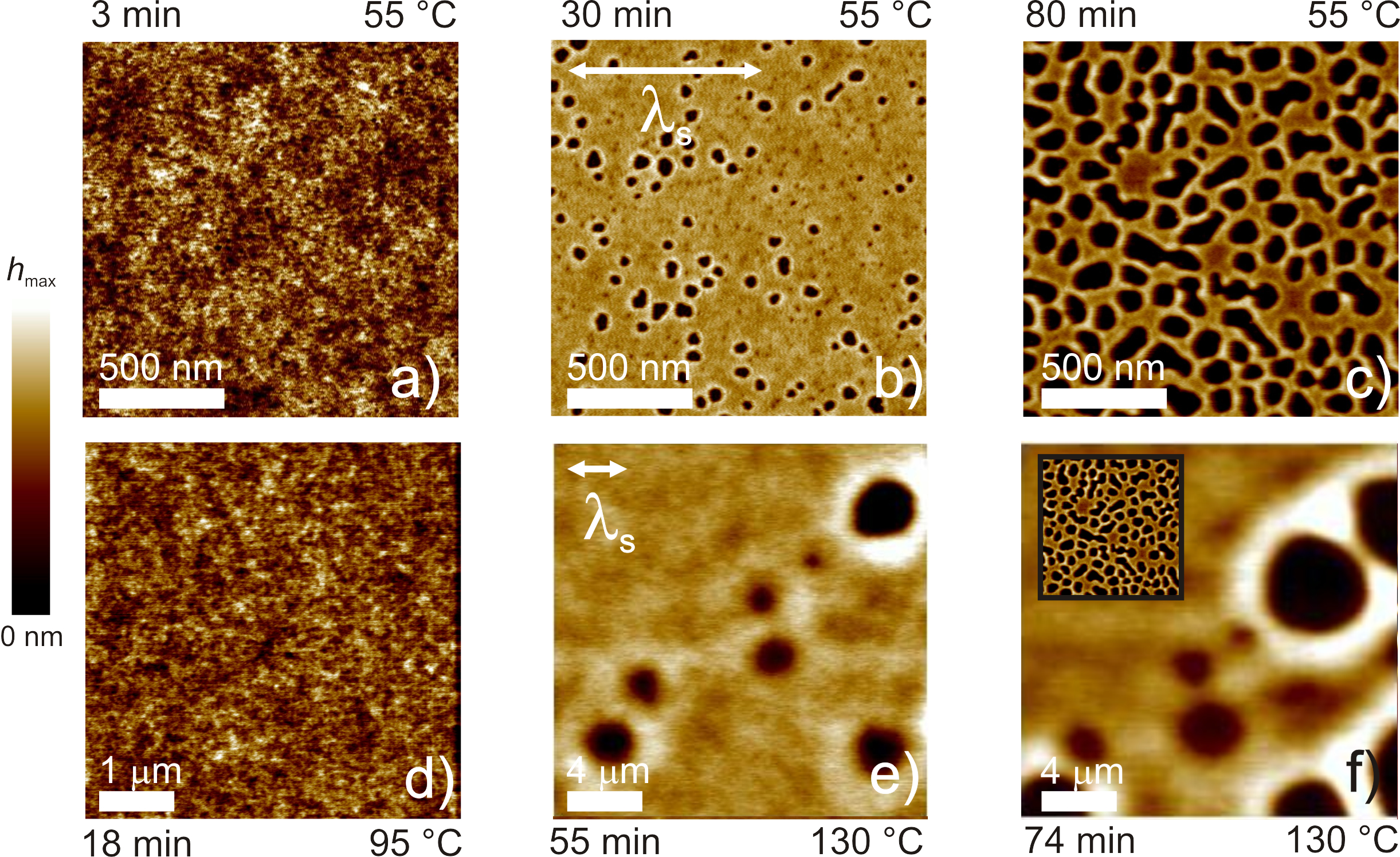}
\end{minipage}%

\vspace{1mm}
\begin{minipage}{\columnwidth}
	 \includegraphics[width=\columnwidth]{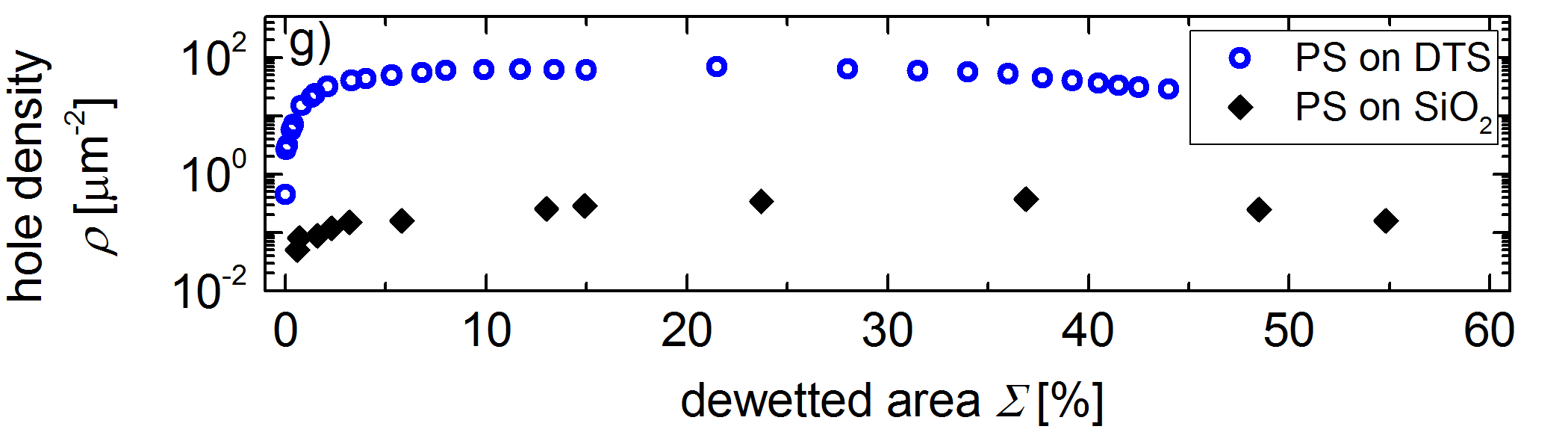}
\end{minipage}%
\caption{a) - f) AFM topography images of ultra-thin dewetting polymer films, on DTS with undisturbed film thickness $h_0 = 5.3(3)$\,nm (top row) and on SiO$_2$ with $h_0 = 6.0(3)$\,nm (bottom row). From a) to c) and d) to f), the height scales, $h_\textrm{max}$, are 2, 6 and 10\,nm. The spinodal wavelengths, $\lambda_\textrm{s} \approx 800$\,nm, predicted from~\cite{Seemann2001, Israelachvili1972} are shown in the middle column for both substrates. The inset of f) shows c) at the same scale. g) A comparison of the hole density $\rho$ as a function of dewetted area $\Sigma$ for PS films dewetting from DTS and SiO$_2$ shown in a) - f). }
\label{Fig:Vergleich}
\end{figure}

In this work, we present two systems that are nominally susceptible to spinodal dewetting.  The first is a polymer film on a silicon wafer with thermally oxidzed layer. The second system is an \emph{identical} polymer film on an \emph{identical} wafer, additionally covered with a hydrophobic SAM. Models used to describe the spinodal decomposition on no-slip substrates~\cite{Seemann2001,Becker2003} have been extended to strong slip systems by Rauscher and co-workers~\cite{Rauscher2009}. With regard to the second system studied here, the no-slip theory~\cite{Seemann2001,Becker2003} predicts that spinodal decomposition of thin PS films may occur, even in cases of strong slip. However, our experimental observations are incompatible with this prediction. Rather, the experimental results provide evidence of a random nucleation process. Thus, in the situation of a film that combines characteristics of the free-standing case (\emph{i.e.} large slip), with those of a solid support containing energetics promoting spinodal instability, the actual breakup observed is that of a free-standing film. 

As a dewetting liquid, polystyrene with a molecular weight of 4.2\,kg/mol (PSS, Mainz, Germany) was used. The films with thicknesses $4\textrm{\,nm}\lesssim h_0 \lesssim 7\textrm{\,nm}$ were supported on the two types of substrates. The first exhibits a 150\,nm SiO$_2$ layer at the top of a Si wafer (Si-Mat, Kaufering, Germany), which we refer to as the SiO$_2$ substrate. To change the solid/liquid boundary condition as well as the surface energy, the second type additionally incorporated a SAM composed of alkylsilane molecules (dodecyltrichlorosilane, DTS, Sigma-Aldrich, Darmstadt, Germany) with $b \approx 500$\,nm $\gg h_0$~\cite{Fetzer2005,Fetzer2007}\footnote{The quoted $b$ was obtained from nucleated hole growth dewetting experiments of 5.6\,kg/mol PS films with $h_0\approx$\,100\,nm (at $120\,^\circ$C). This measurement of $b$ was not performed with the ultra-thin films discussed here, since the models used~\cite{Fetzer2005,Fetzer2007} to extract $b$ are not valid when vdW forces play a major role as with the films studied here. }. As described in detail elsewhere~\cite{Lessel2012}, the RMS surface roughness of a DTS coated SiO$_2$-substrate is identical to a bare one (approx.\ 0.2\,nm), while providing a surface energy $\gamma_\textrm{DTS} = 26(1)$\ mJ/m$^2$ and leading to an equilibrium contact angle $\theta_\textrm{Y} = 66(2)\,^\circ$ for PS. For details see Ref.~\cite{Lessel2012}. PS was spincast from toluene (chromatography grade, Merck, Darmstadt, Germany) onto a mica sheet (V2 Grade, PLANO GmbH, Wetzlar, Germany) and then annealed at 80\,$^{\circ}$C ($> T_g \approx 75\,^{\circ}$C~\cite{santangelo98MAC}) for 1\,h to release residual stress and solvent. After annealing, the films were floated onto an ultra pure water bath (18.2 M$\Omega$\,cm, total carbon content $<$\ 6\ ppb, TKA-GenPure, TKA Wasseraufbereitungssysteme GmbH, Braunschweig, Germany), and then transferred to the substrates. By mounting several wafers on a glass slide, it was possible to cover them simultaneously from one PS film, ensuring identical film thickness on multiple samples.

The experiments were performed using \emph{in situ} high speed atomic force microscopy with scan areas between $1.5\times1.5\,\upmu$m$^2$ and $20\times20\,\upmu$m$^2$ (AFM, Dimension FastScan featuring a Dimension heater with Fastscan-A tips, Bruker, Karlsruhe, Germany)~\footnote{Influence of the AFM scanning on the dewetting process was excluded by periodically enlarging the scan area and scanning independent areas after a quench; no difference in morphology was observed. }. While a detailed study of the temperature dependence is not the subject of this article, we note that in all cases the ultra-thin films on DTS dewetted at temperatures below bulk $T_g$; experimentation at these low temperatures, $55^\circ\textrm{C} \leq T \leq 65\,^\circ\textrm{C}$ which are significantly less than bulk $T_\textrm{g}$, was required to enable observation of the dewetting on accessible timescales (tens of minutes). These observations are consistent with dewetting of ultra-thin films on SiO$_2$ substrates observed previously~\cite{Seemann2001, Herminhaus2004,yang10SCI, Xu2014}.

A comparison between the dewetting process of PS films on a hydrophobic DTS substrate and a hydrophilic SiO$_2$ substrate is depicted in Fig.~\ref{Fig:Vergleich}. The films exhibit an almost identical film thickness (on SiO$_2$: 6.0(3)\,nm, on DTS: 5.3(3)\,nm). Effective interface potentials can be computed according to the layer model developed in Refs.~\cite{Israelachvili1972,Seemann2001} and applied to the DTS coated or uncoated SiO$_2$ substrates. The effective interface potential is approximated~\cite{Seemann2001, blossey2012, dietrich13TXT, Oron1997} by 
\begin{equation}
\phi(h) = C/h^{8} -A/(12\pi h^{2})\ .
\label{eip}
\end{equation}
The Hamaker coefficient, $A = 2.05\times10^{-20}$\,J, is taken as that for the air/PS/SiO$_2$~\footnote{The buried Si substrate does not need to be included since the SiO$_2$ layer is 150 nm thick, which is much longer than the range of the vdW interactions considered here.} system for both cases since the optical index of refraction of DTS is nearly identical to that of SiO$_2$~\cite{Seemann2005, wasserman89JACS, isf11TXT, Lessel2012}. $C$ is chosen to obtain the appropriate contact angle through $\phi(h_\textrm{min}) = \sigma(1-\cos\theta_\textrm{Y})$~\cite{Seemann2001, Seemann2005,degennes03TXT}, where $h_\textrm{min}$ is the film thickness at which $\phi$ reaches a global minimum and $\sigma$ is the polymer/air surface tension. While $A$ for SiO$_2$ and DTS are similar, $C$ for the two substrates vary substantially with $C_{\textrm{SiO}_2}
=6.3(1)\times10^{-22}$\,J\,nm$^6$ and $C_\textrm{DTS}=8.2(1)\times10^{-26}$\,J\,nm$^6$. In the absence of slip, the expected spinodal wavelengths on both substrates are at least 800 nm, which were calculated as in Ref.~\cite{Seemann2001} (see Supplementary Material Figs. S1 and S2). As shown in Fig.~\ref{Fig:Vergleich}, we find that the typical hole-to-hole distances in the PS film on the bare SiO$_2$ substrate (lower row) are consistent with this prediction; further evidence for spinodal dewetting as the mechanism for dewetting from SiO$_2$ is provided in the Supplementary Material, Figs.~S3 and~S4. The spatial structures in the PS film on DTS (upper row), however, are incompatible with the this spinodal breakup scenario, since the distance between hole nuclei is much less than the expected distance. 
\begin{figure}[t!]
\centering
\includegraphics[width=0.98\columnwidth]{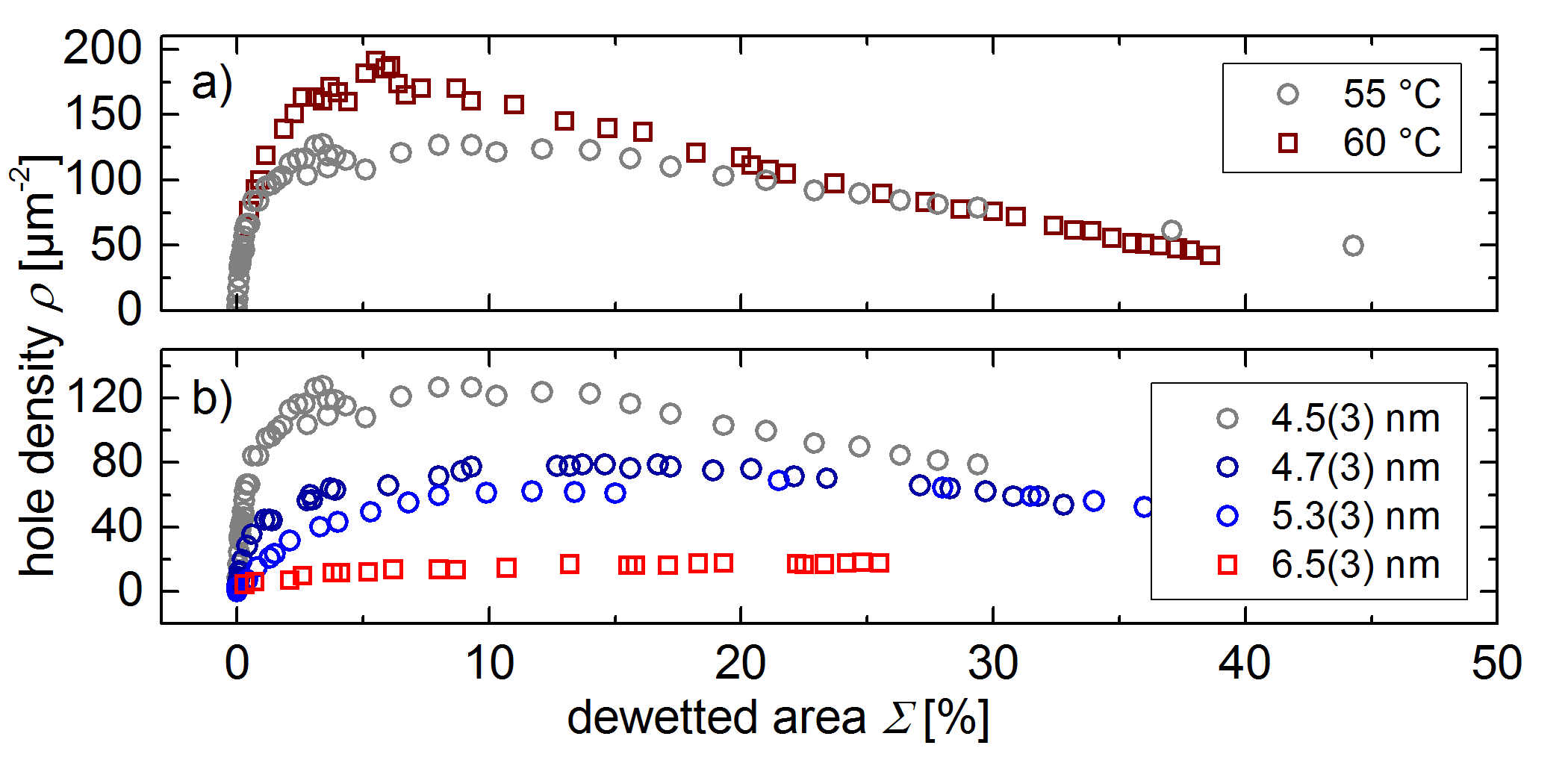}
\caption{Hole densities observed on DTS as a function of dewetted area for a) different dewetting temperatures and thickness of 4.5(3) nm and b) films of different thickness; all dewetted at 55\,$^\circ$C, except for the thickest film for which it was necessary to increase the temperature to 65\,$^\circ$C in order to observe dewetting. }
\label{Fig:Results}
\end{figure}

Hole densities $\rho = N/L^2$ with $N$ the number of holes observed in AFM images (scan area $L^2$) for both types of substrates are compared in Fig.~\ref{Fig:Vergleich}g). The hole densities are plotted as a function of the fractional dewetted area $\Sigma$, thereby taking into account the enhanced dewetting dynamics on DTS due to the presence of slip. As such, effects of different temperatures and boundary conditions can be excluded~\cite{Nguyen2014}. At low hole radii and for early times, hole densities increase as nucleation events occur. After nucleation, hole growth proceeds and the dewetted area increases while holes continue to nucleate in undewetted areas.  At large hole radii, coalescence of holes begins and the densities decrease after reaching a maximum. As described above, the typical hole-to-hole distance as depicted in Fig.~\ref{Fig:Vergleich} is consistent with the spinodal wavelength inferred from the effective interface potential~\cite{Seemann2001, Becker2003}. Rauscher \emph{et al.}~\cite{Rauscher2009} predicted that the presence of strong slip ($b\gg h_0$) would lead to an enlargement of the expected spinodal wavelength, resulting in a decreased hole density. However, the hole densities observed on the DTS substrates are three orders of magnitude higher than those seen on SiO$_2$. 

A higher hole density (that is, a smaller typical hole-to-hole distance) does not necessarily exclude a spinodal decomposition of the film. Since holes that originate from a spinodal dewetting process are correlated~\cite{Jacobs1998,Herminghaus1998}, spinodal dewetting is not compatible with a random distribution of holes. In order to check for correlations between objects in a space, Minkowski functional analysis~\cite{arns04CSA, Mecke2005,Fetzer2007b} has been applied. The Minkowski functionals, which are shown for both SiO$_2$ and DTS in the Supplementary Material, unambiguously exclude spinodal dewetting as the underlying rupture mechanism, and instead prove that the breakup mechanism for films on DTS is a random nucleation process. 

\begin{figure}[b!]
\centering
\includegraphics[width=\columnwidth]{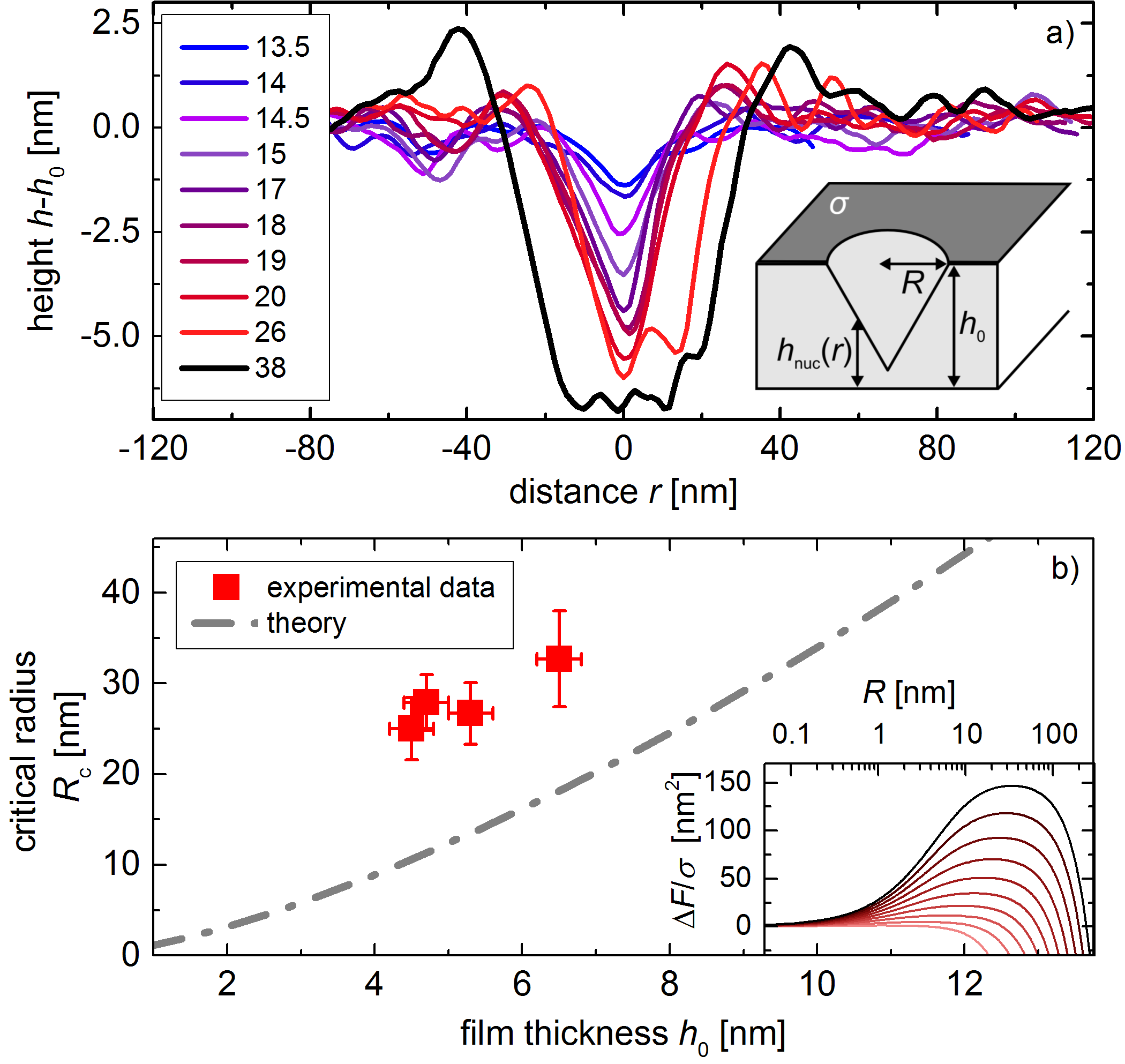}
\caption{a) \emph{In situ} observation of the nucleation of a hole in a PS film with $h_0 = 5.2(3)$\,nm on DTS (\emph{cf.} Fig.~\ref{Fig:Vergleich}). The legend indicates the time, in minutes, of annealing at 55\,$^\circ$C. b) Critical radii for the films represented in Fig.~\ref{Fig:Results}b); vertical errors represent the statistical distribution of $R_\textrm{c}$ measured in the experiments. The insets show a) schematic depiction of a hole nucleus as modelled in the text and b) predictions for $\Delta F/\sigma$ ($1<h_0<10$\,nm, increment 1 nm), from which critical radii, $R_c$, are obtained. }
\label{Fig:Model}
\end{figure}

To explore the nature of the hole nucleation mechanism of ultra-thin liquid films dewetting from DTS, the influence of the film thickness and dewetting temperature were investigated. In Fig.~\ref{Fig:Results}a) the hole density is plotted as a function of dewetted area for films of identical thickness, but different temperatures; an increase in temperature leads to higher maximum hole density. Fig.~\ref{Fig:Results}b) shows that a thinner film also leads to a higher maximum hole density. In the context of a random nucleation mechanism, both a higher temperature and thinner film are expected to lead to a higher probability of a nucleation event. 

In Fig.~\ref{Fig:Model}a) we show AFM height profiles following the nucleation of a hole in a PS film on DTS. At about 20 minutes of annealing, the air/liquid interface makes contact with the substrate, at which time we consider the nucleation event to have taken place with a nucleus (that is, an interface profile, $h_\textrm{nuc}(r)$) in the shape of a cone. Consistent with the ensuing flow following nucleation, this observed cone geometry in Fig.~\ref{Fig:Model}a) corresponds with a contact angle $\arctan(h_0/R) = 14\,^\circ \ll \theta_\textrm{Y} = 66(2)\,^\circ$~\cite{Lessel2012}; a lower than equilibrium contact angle leads to a receding contact line and the hole begins to grow. We also note the buildup of a monotonically decaying rim (\emph{cf.} Fig.~\ref{Fig:Model}a) at 38 min) which is a signature of a strong slip boundary condition~\cite{Fetzer2005, Baeumchen2009a}, \emph{i.e.} $b \gg h_0$. The monotonicity of the rim precedes the formation of satellite holes~\cite{Becker2003} which would be observed for no-slip dewetting characterized by oscillatory height profiles~\cite{Fetzer2005, Baeumchen2009a}. Thus, the slip boundary condition is a necessary condition to observe these morphologies reminiscent of unsupported films. 

To model the hole nucleation process, we assume an idealized conical profile: $h_\textrm{nuc}(r) = h_\textrm{min}+rR^{-1}(h_0-h_\textrm{min})$. Here, $r$ is the radial coordinate and $R$ is the radius of the hole at the undisturbed film thickness. The inset of Fig.~\ref{Fig:Model}a) shows a schematic representation and variable definitions. The excess free energy of such a cone can be written as the sum of two contributions: 
\begin{align}
\Delta F[\phi(h),h]=\Delta F_\textrm{surf}+\Delta F_\textrm{vdW}\,,
\label{Eq:free}
\end{align}
where the excess area
\begin{equation} 
\Delta F_\textrm{surf}/\sigma = \pi R \sqrt{R^2+(h_0-h_\textrm{min})^2}-\pi R^2\ ,
\end{equation}
is the surface contribution, and the surface tension is $\sigma = 0.038$\,J/m$^2$~\cite{wu70JPC} at the temperatures investigated here. Since the films are only some nanometers thick, a vdW contribution is necessary and is expressed as 
\begin{align}
\Delta F_\textrm{vdW} = 2\pi\int_0^R\,\textrm{d}r\,r\phi(h_\textrm{nuc}(r))-\pi R^2\phi(h_0)\,. 
\label{vdwdef}
\end{align}
With $\phi$ defined as in Eq.~\ref{eip}, we find that 
$\Delta F_\textrm{vdW} = -c_\textrm{vdW}R^2$,
where $c_\textrm{vdW}(h_0, C, A)$ is a positive constant depending on the details of the profile $h_\textrm{nuc}(r)$ and effective interface potential. Here we use $C_\textrm{DTS}$ and $A$ defined below Eq.~\ref{eip}. 

In general, $\Delta F_\textrm{surf}$ is positive, obtaining a plateau for $R/(h_0-h_\textrm{min})\gg 1$, while $\Delta F_\textrm{vdW}$ is, for the monotonically increasing $\phi(h)$ presented here, negative (\emph{n.b.} $h \geq h_\textrm{min}$). As such, the free energy model of Eq.~\ref{Eq:free} predicts free energy profiles, shown in the inset of Fig.~\ref{Fig:Model} for $1\leq h_0 \leq 10$\ nm; plots of the individual components are found in the Supplementary Material. The global maxima define a critical radius, $R_\textrm{c}$, beyond which the cone grows. $R_\textrm{c}$ can also be obtained from the the observation of hole formation in the AFM measurements (\emph{e.g.} at 20\,min in Fig.~\ref{Fig:Model}a)). These $R_\textrm{c}$ obtained from at least 25 holes for a given film, are shown in Fig.~\ref{Fig:Model}b). The experimental data is consistent with the predictions of the model developed from Eq.~\ref{Eq:free} with no free parameters. 

While the experimental data and model presented in Fig.~\ref{Fig:Model}b) are consistent, the associated energy barriers are larger than the available thermal energy ($kT/\sigma \approx 0.1$\,nm$^2 <\Delta F(h_0=5\,\textrm{nm}, R_\textrm{c})/\sigma \approx 35$\,nm$^2$), yet are comparable to the nuclei energies predicted for unsupported films~\cite{Croll2010,Rathfon2011}. We may thus consider certain heterogeneities to aid in the nucleation process. For heterogeneous nucleation due to dust contamination, holes typically appear with a monodisperse size distribution~\cite{Seemann2001, Seemann2005}; by contrast, Fig.~\ref{Fig:Vergleich}b) shows PS dewetting from DTS with a broad distribution of hole sizes and Figs.~\ref{Fig:Results}a) and b) show a continuous appearance of holes. Furthermore, in previous supported film studies~\cite{Seemann2001, Seemann2005}, in a majority of the hole centers a nucleus could be clearly identified. Yet for dust-free thin films with typical nanometric dimensions as studied here and elsewhere~\cite{Croll2010,Rathfon2011}, no nuclei in the form of dust particles can be observed in either experiment (optical resolution in refs.~\cite{Croll2010,Rathfon2011}, and pixel sizes as small as 1.5\,nm in the AFM measurements here). Here we propose as in~\cite{Jacobs1998} that heterogeneities in the form of free volume (voids), on the scale of several angstroms~\cite{liuMAC93, yuJPSB94, banJPSB96}, may overlap in rare instances to aid in the nucleation of a hole. A detailed statistical analysis of such events is beyond the scope of the present article, yet we hope that future efforts may elucidate the possible role of such processes. Regardless of the important distinction between heterogeneous and homogeneous nucleation, we reiterate that the same film on two different substrates (\emph{i.e.} SiO$_2$ and DTS) breaks up on different time scales and with a completely different mechanism, despite their similar energetics at the undisturbed film thicknesses. 

In conclusion, we have studied dewetting polystyrene films with thicknesses $4\ \textrm{nm} \lesssim h_0 \lesssim 7$\,nm on hydrophilic and hydrophobic substrates. In agreement with previous studies, dewetting from the hydrophilic substrates proceeds via spinodal decomposition, \emph{i.e.}\ through the amplification of capillary waves. The lower-energy hydrophobic substrates provide a slip boundary condition at the solid/liquid interface, thus bridging the gap between supported film breakup and the bursting of unsupported films. Consistent with previous findings, but here for hydrophobic substrates, we observe a sub-$T_\textrm{g}$ dynamics that is much faster than observed for no-slip substrates. The governing breakup mechanism is also different. We provide evidence that the films, which are nominally susceptible to spinodal decomposition, breakup through a nucleation mechanism if supported by slippery hydrophobic substrates, which is instead consistent with the behaviour observed in unsupported films providing a full slip boundary condition and no substrate surface energy.

The authors acknowledge financial support from the German Science Foundation (DFG), the Graduate College 1276 and NSERC of Canada. OB acknowledges support from the ESPCI Joliot Chair. JDM was supported by LabEX ENS-ICFP: 
ANR-10-LABX-0010/ANR-10-IDEX-0001-02 PSL. The authors also thank Sabrina Haefner, Mischa Klos, Markus Rauscher and Kari Dalnoki-Veress for insightful discussions. 

\providecommand{\noopsort}[1]{}\providecommand{\singleletter}[1]{#1}%

\clearpage

\renewcommand{\theequation}{S\arabic{equation}}
\renewcommand{\thefigure}{S\arabic{figure}}
\setcounter{figure}{0}

\section*{Supplementary Material for: Nucleated dewetting in supported ultra-thin liquid films with hydrodynamic slip}

\subsection*{Effective Interface Potential and Spinodal Wavelength}

\begin{figure}[h!]
\centering
\begin{minipage}{0.47\textwidth}
\includegraphics[width=0.98\textwidth]{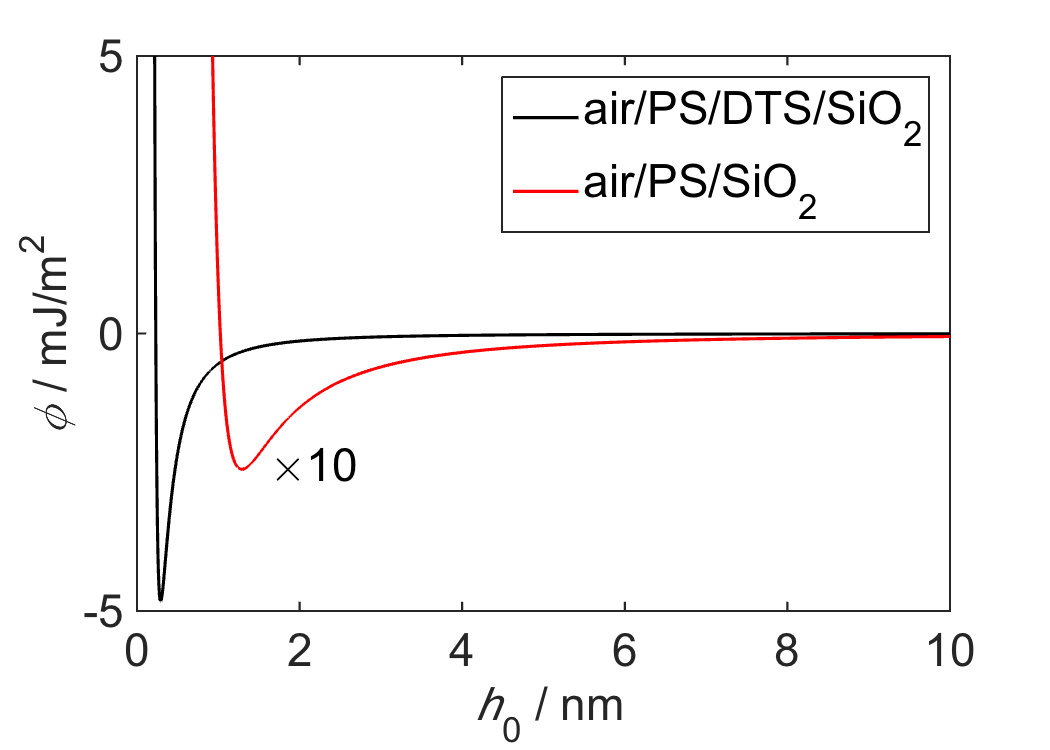}
\caption{Effective interface potentials, $\phi$, for PS on a DTS coated SiO$_2$ wafer and for PS on a SiO$_2$ wafer. For clarity $\phi$ for the latter has been multiplied by ten. We have made use of $C_\textrm{DTS} = 8.2\times 10^{-26}$\,J/nm$^6$, $C_{\textrm{SiO}_2} = 6.3\times 10^{-22}$\,J/nm$^6$, and $A = 2.05\times 10^{-20}$\,J, as well as Eq. 1 of the main text. }
\label{Fig:effints}
\end{minipage}%
\quad
\begin{minipage}{0.47\textwidth}
\includegraphics[width=0.98\textwidth]{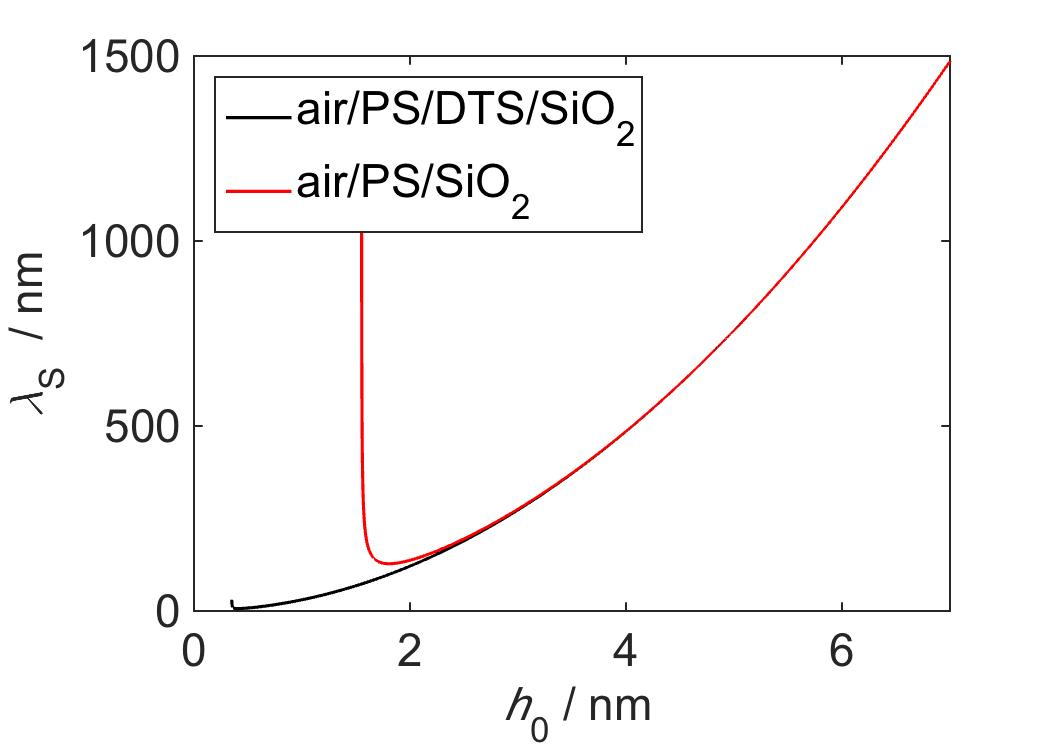}
\caption{Spinodal wavelengths predicted using $\lambda_\textrm{S} = \sqrt{-8\pi^2\sigma/\phi''}$ and the effective interface potentials of Fig.~\ref{Fig:effints} and $\sigma =  0.038$\,J/m$^2$. We note that the predicted spinodal wavelengths do not significantly differ for $h_0 > 3$\,nm for the two cases considered here. }
\label{Fig:lambdas}
\end{minipage}%
\end{figure}

\subsection{Minkowski Measures; Power Spectral Densities}
Different methods are suitable to analyze correlations in point patterns, in our case, Minkowski measures have shown to be extremely convenient, since they are able to find not only two-point correlations (like a Fourier transformation oder a pair correlation function g(r)), but also capture higher-order correlations~\cite{mecke98IJMP}. In the very same way as in earlier studies, we have analyzed the statistical distribution of the hole sites by Minkowski measures~\cite{Jacobs1998,Herminghaus1998}. In two dimensions (like for our point set of hole sites), three Minkowski measures completely describe the distribution and there are analytical descriptions of what to expect for a random (Poissonian) set of sites. This is the clear strength of Minkowski measures, a random process can unambiguously be detected analytically. Deviations of that expectation then show that the hole sites are correlated. 
\begin{figure}[t!]
\centering
\begin{minipage}{0.47\textwidth}
\includegraphics[width=0.98\textwidth]{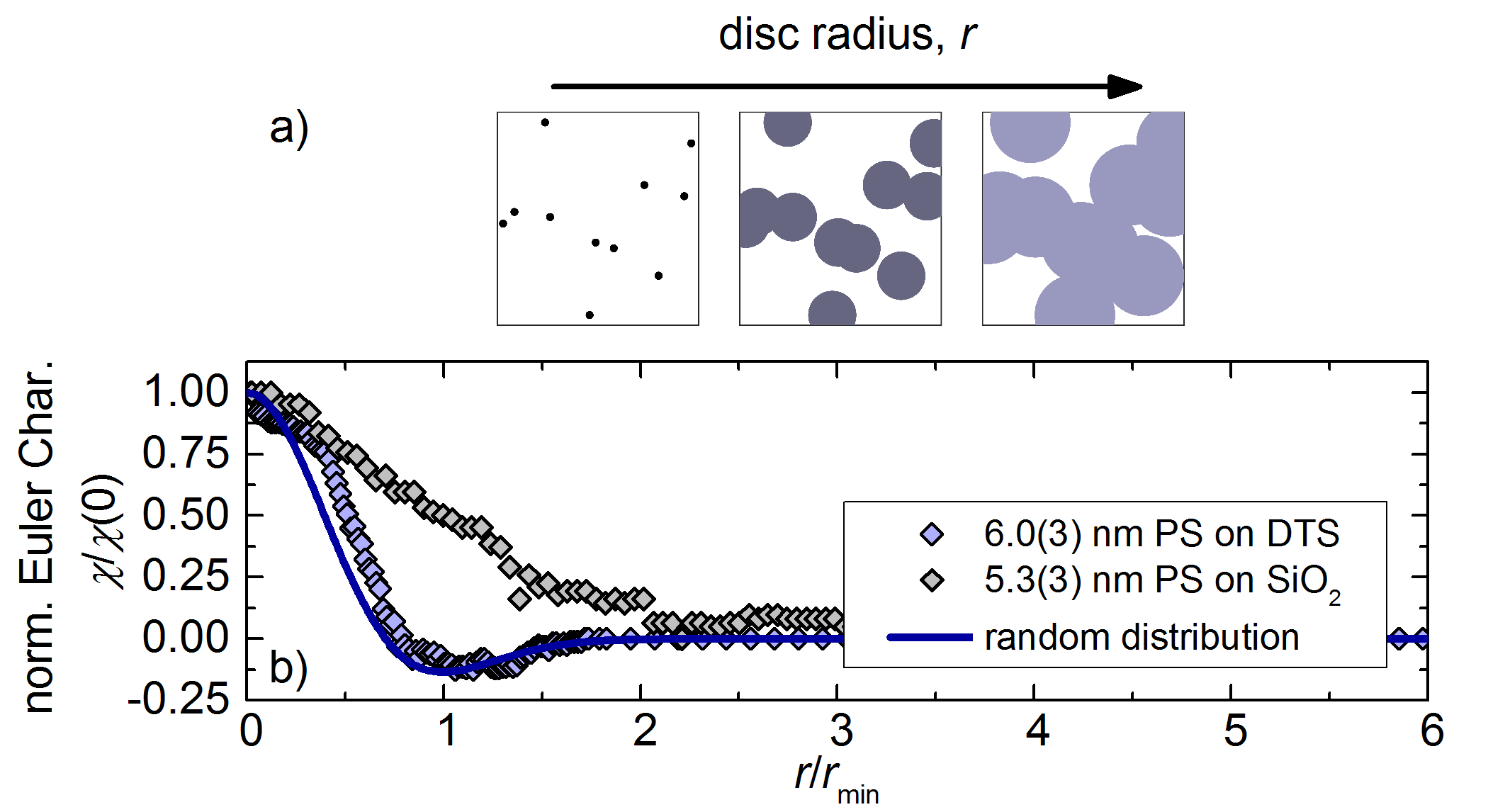}
\caption{a) Schematic depiction of the procedure for determining the Euler characteristic, $\chi$: hole centers are identified, and disks of variable radius $r$ are placed at each center. As the radius of each circle grows, fewer and fewer independent objects cover the plane, thus reducing $\chi$. b) Statistical distribution of holes revealed by the normalized $\chi$, for films at a similar dewetted area (symbols), and the theoretical prediction for uncorrelated holes (line) with no fitting parameter. }
\label{Fig:Mink}
\end{minipage}%
\quad
\begin{minipage}{0.47\textwidth}
\includegraphics[width=0.98\textwidth]{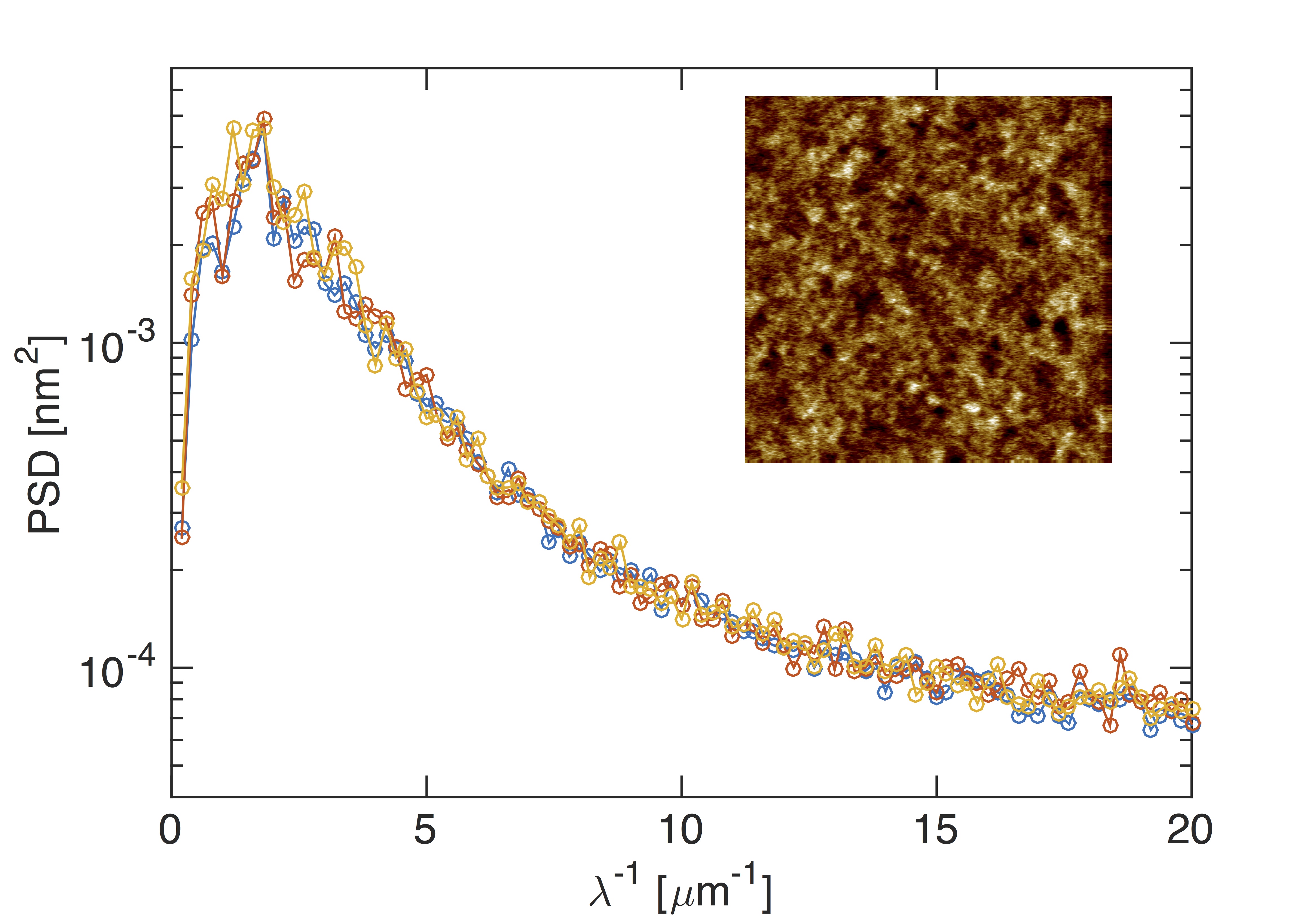}
\caption{Power spectral densities (PSD) of three sequential \emph{in situ} AFM images (inset, horizontal scan size $5\times5\,\mu$m$^2$, vertical scale 1.8 nm), for a 6 nm PS film dewetting from SiO$_2$, taken at 115\,$^\circ$C with $36\,\textrm{min} \leq t \leq 43\,\textrm{min}$. The positions of the maxima of the three curves coincide and correspond very well with the expected spinodal wavelength predicted using the effective interface potential shown in Fig.~\ref{Fig:effints}. }
\label{Fig:PSD}
\end{minipage}%
\end{figure}

One Minkowski measure is the Euler characteristic, $\chi$, which quantifies the connectivity of a distribution of objects -- in this case the distribution and connectivity of the holes seen in Fig.~1. 
$\chi$ is computed by placing disks of radius $r$ at the center of every hole (see Fig.~\ref{Fig:Mink}a)), and is thereby only a function of the number of holes and the disk radius, $\chi = \chi(r,N)$. It is also possible to analytically calculate $\chi$ for a Poisson distribution of holes, as described in Refs.~\cite{Stoyan1996,arns04CSA,Mecke2005}. Fig.~\ref{Fig:Mink}b) depicts the Euler characteristic of the experiments shown on Fig.~1 
 and that obtained by the theory for a Poisson distribution (eq. 24 in~\cite{arns04CSA}): 
 \begin{equation}
 	\chi(r) = \pi\rho(1-\pi r^2\rho)e^{-\pi r^2\rho}\ ,
 \end{equation}
where $\rho$ is the number density of disks, easily accessed from the experimental data. The prediction of the Poisson distribution, a universial curve in the normalized representation of Fig.~\ref{Fig:Mink}b), is not followed by the hole formation of spinodally dewetting films, as seen in~\cite{Becker2003, Herminghaus1998} and confirmed here for the liquid films on the bare SiO$_2$ substrate; the power spectral densities of Fig.~\ref{Fig:PSD} show furthermore a well defined peak at the expected spinodal wavelength, giving further evidence for the spinodal mechanism governing the film breakup on SiO$_2$ substrates. In contrast, the Euler characteristic for dewetting films on DTS is well captured by the uncorrelated random distribution, which proves that the dewetting process on DTS cannot originate from a spinodal process.

\subsection{Free Energy Components}

\begin{figure}
\centering
\begin{minipage}{0.47\textwidth}
\includegraphics[width=0.98\textwidth]{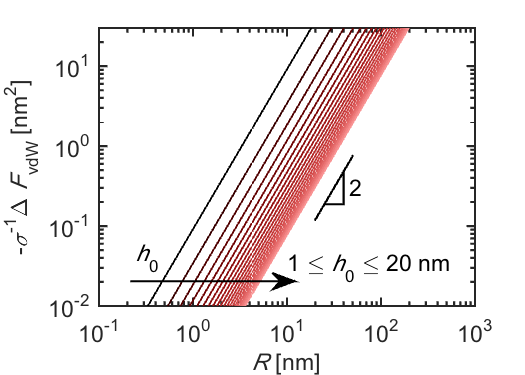}
\caption{Representation of the van der Waals contribution to the free energy, $\Delta F_\textrm{vdw}$ as computed from Eqs. (1) and (4). The slope of the curves in this log-log representation confirms the relation $\Delta F_\textrm{vdW} = -c_\textrm{vdW}R^2$. \emph{n.b.} the minus sign on the vertical axis allowing for a log representation. The arrow points in the direction of increasing $h_0$. }
\label{Fig:vdwconts1}
\end{minipage}%
\quad
\begin{minipage}{0.47\textwidth}
\includegraphics[width=0.98\textwidth]{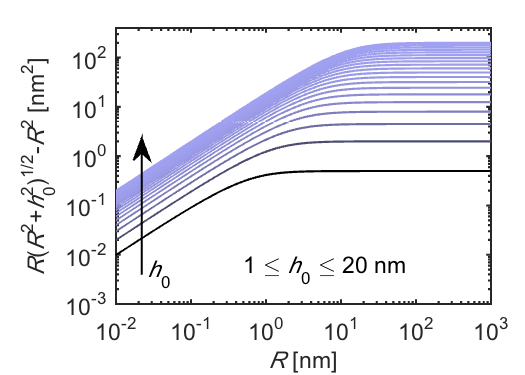}
\caption{Representation of the capillary contribution to the free energy, $\Delta F_\textrm{surf}/(\pi\sigma)$ in Eqs. (2) and (3).  The arrow points in the direction of increasing $h_0$. }
\label{Fig:vdwconts2}
\end{minipage}%
\end{figure}

\providecommand{\noopsort}[1]{}\providecommand{\singleletter}[1]{#1}%

\end{document}